\documentclass[a4paper,11pt]{article}
\pdfoutput=1 

\usepackage{jheppub} 

\usepackage{mathpazo} 
\usepackage{amsmath}
\usepackage{amssymb}
\usepackage{hyperref}
\usepackage{float}
\usepackage{url}
\usepackage{mathtools}
\usepackage[T1]{fontenc} 

\title{\boldmath Hadron energy estimation from Atmospheric neutrino events}

\author[a,b,1]{M. Nizam,\note{Corresponding author.}}
\author[c,2]{S. Uma Sankar\note{Corresponding author.}}

\affiliation[a]{Tata Institute of Fundamental Research,\\Mumbai 400005, India}
\affiliation[b]{Homi Bhabha National Institute,\\Mumbai 400094, India}
\affiliation[c]{Department of Physics, Indian Institute of Technology Bombay,\\Mumbai 400076, India}

\emailAdd{mohammad.nizam@tifr.res.in}
\emailAdd{uma@phy.iitb.ac.in}

\abstract{The ICAL at INO is designed to mainly observe the  muons produced in the charged current interactions of atmospheric muon neutrinos and anti-neutrinos. The track of the muon is reconstructed using the hits they produce in the detector. From this track, the charge, the energy and the direction of the muon are estimated, which are used to do oscillation physics analysis. In a large fraction of events, a number of hadrons are also produced in addition to the muons. The charged hadrons also leave hits in the detector which can be utilized to estimate the hadron energy. In this work, we study the relation between hadron hits, defined to be the difference between the total number of hits and the muon track hits, and the hadron energy. We find that a non-negligible number of baryons are produced in atmospheric neutrino interactions. For $E_{had} < 5$ GeV, almost all the hadron energy is carried by these baryons. Finally, we formulate a procedure by which the hadron energy can be estimated from the number of hadron hits.}

\begin{document} 
\maketitle
\flushbottom

\section{Introduction}
\label{sec:intro}

The Iron calorimeter (ICAL) at the India-based neutrino observatory (INO) is a giant magnetized neutrino detector. Its primary goal is to study the interactions of the atmospheric muon neutrinos and anti-neutrinos. It aims to determine neutrino mass hierarchy and also add to the precision of neutrino oscillation parameters \cite{a}.

ICAL has a mass of 50 ktons and it consists of three modules, each of which has dimensions 16 m $\times$ 16 m $\times$ 14.4 m. Each module contains 151 layers of 5.6 cm thick  iron plates interspersed with resistive plate chambers (RPCs). 
The iron plates act as targets to atmospheric neutrinos. The charged current (CC) interactions of $\nu_{\mu}$ and $\bar{\nu}_{\mu}$ produce muons. Because of the 1.3~T magnetic field in the iron plates, ICAL can determine the charge of the muon and distinguish between the interactions of $\nu_{\mu}$ and $\bar{\nu}_{\mu}$. There are approximately 30,000 RPCs in the detector. Each RPC has a surface area of 1.84 m $\times$ 1.84 m and carries a potential difference of 10 kV across its electrodes. Copper strips, of width 1.96 cm, are laid in parallel on the top (bottom) surface of the RPC in X (Y) direction. The vertical direction is taken to be the Z-direction. A charge particle, passing through the RPC, creates an avalanche. The signal due to this avalanche is read by the X and Y copper strips. These readings, together with the RPC layer number, give (X,Y,Z) coordinates of a "hit".

The muons produced in the detector, being minimum ionising particles, pass through a number of iron layers and RPCs. Based on the hits produced by the muon in different RPCs, it is possible to reconstruct the track of the muon. The hits which make up the track are defined to be "track hits".  
From the reconstructed track, the charge, the energy and the direction of the muon are estimated with good precision. In the first set of physics studies, the physics capabilities of ICAL were estimated using the kinematical information of only the muons \cite{b,c,d}.

In addition to a muon, an atmospheric neutrino interaction also produces a set of hadrons (mesons and baryons). The energy of these hadrons can be estimated based on the hits produced by the charged hadrons in the detector. However, we need to develop a different technique for this estimation. Most of the time, the hits due to the hadrons can not be reconstructed into tracks because (a) the energy of a typical hadron is much smaller than the energy of the muon and (b) the hadrons can be absorbed by the detector nuclei. Thus, the inclusion of the hadron energy in the kinematic reconstruction of an event poses a great challenge. 
The first estimate of the hadron energy of atmospheric neutrino events in ICAL was done in ref. \cite{e}. 

Various efforts were made to estimate the hadron energy in ICAL \cite{f,g}. However, in these efforts, a charged pion of known energy is injected into Geant4 simulator and the corresponding hit pattern was studied. This process was repeated for pions of different energies and different directions with respect to the zenith. Through these simulations, a correlation between the pion energy and the number of hits was established and the resolution in pion energy was estimated. \textit{It was assumed that these correlations and the resolutions will hold for all hadrons produced in atmospheric neutrino interactions}.  The physics capabilities of ICAL are re-calculated using three kinematical variables: muon energy, muon direction and the hadron energy, estimated from the pion simualtions mentioned above. With this "3-D analysis", it was shown that the physics capabilities are enhanced \cite{h,i}.

In this work, we did a systematic study of particle production in atmospheric neutrino events. We found that a significant number of baryons are produced in a large fraction of these events. For hadron energy less than 5 GeV, these baryons carry almost all the hadron energy. Therefore, we believe that the hit pattern produced by an isolated, single charged pion does not represent the hit pattern produced by the hadrons in an atmospheric neutrino event properly. We establish a correlation between the hits produced by the baryons and the baryon energy in atmospheric neutrino events and show that this correlation is very different from the correlation found for pions in reference \cite{g}. We then obtain a relation between the number of hits produced by all the hadrons in an
atmospheric neutrino event and the hadron energy. As mentioned before, such a study was done before in ref. \cite{e}. However, we include two features
which were not present in ref. \cite{e}.

To obtain a realistic correlation between the hadron energy in atmospheric neutrino events and the number of hits in ICAL, we used the following procedure: We generated 100 years of unoscillated atmospheric neutrino events through the NUANCE event generator \cite{j}. We did a full Geant4 simulation of all the $\nu_{\mu}$-CC and $\bar{\nu}_{\mu}$-CC events. We selected those events for which the muon track is reconstructed. For these events, we isolated the hits produced by the hadrons by eliminating the muon track hits from the total number of hits. We used this hadron hit bank information to estimate the hadron energy and the energy resolution.

In the analysis below, we use only the total number of hits in an event but not the detailed information of the coordinates of each hit. This helps us in avoiding problems related to "ghost hits" \cite{f}. Ghost hits arise in situations where two or more charged hadrons pass through different regions of the same RPC. If one hadron passes through the point ($X_1, Y_1$) and the other through ($X_2, Y_2$), then we get signals from the two "X" strips at $X_1$ and $X_2$ and from the two "Y" strips $Y_1$ and $Y_2$. Such a signal could have arisen due to the particles passing through the points ($X_1,Y_2$) and ($X_2, Y_1$). Such points are referred to as "ghost hit" points. In our analysis, we avoid this problem by taking the number of hits in an RPC to be the maximum of (number of X strips with a signal, number of Y strips with a signal).
\section{Baryons in atmospheric neutrino events}
\begin{figure}[H]
\centering
\includegraphics[width=15cm, height=11.88cm]{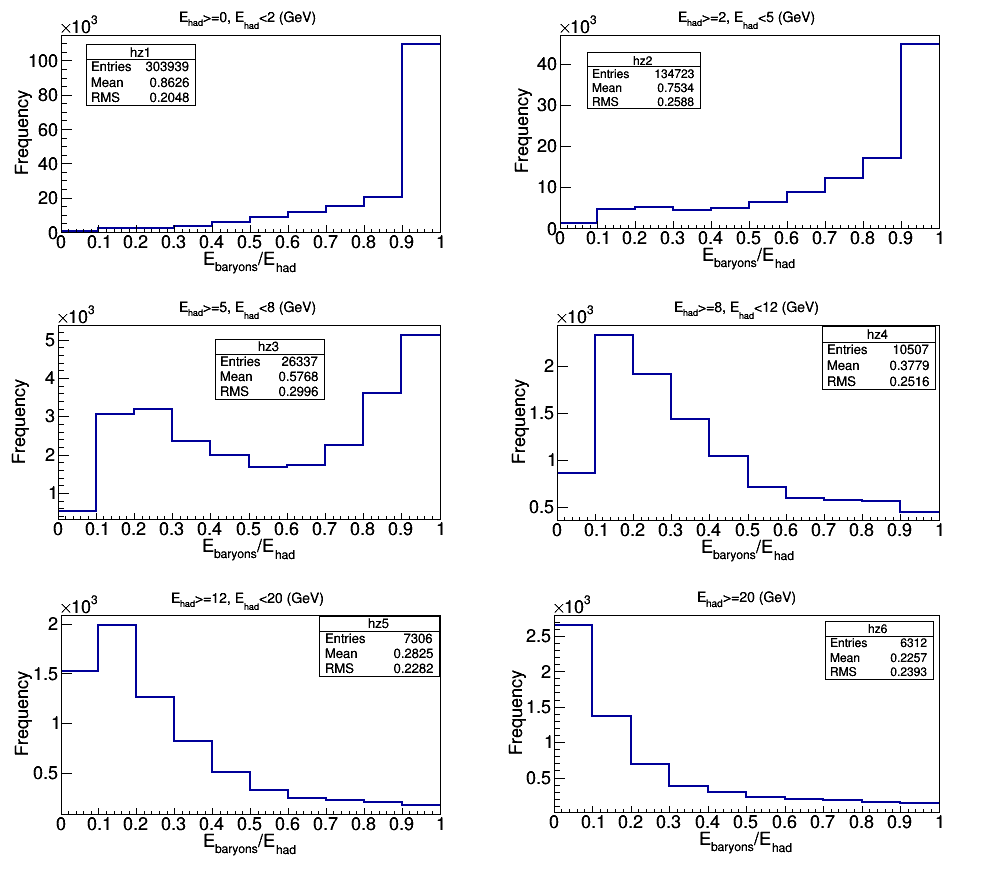}
\caption{$E_{baryons}$/$E_{had}$ vs frequency}
\label{fig:EbaryEhadfrac}
\end{figure}

The analysis presented in this section is also based on 100 years of atmospheric neutrino events generated by NUANCE. As mentioned in the introduction, we believe that doing a simple simulation of isolated charged pions does not reflect the true picture of hadron production in atmospheric neutrino events. This is so because a fair number of baryons are produced in these events and they carry a non negligible fraction of the hadron energy. This is illustrated in figure \ref{fig:EbaryEhadfrac}. In this figure, the two variables $E_{had}$ and $E_{baryons}$ are obtained from NUANCE by the following definitions:
\begin{eqnarray}
E_{had}&=&E_{\nu} - E_{\mu}, \nonumber \\
E_{baryons}&=&E_{\nu} - E_{\mu} - E_{mesons},
\end{eqnarray} 
where $E_{\nu}$ is the energy of the neutrino, $E_{\mu}$ is the energy of the muon and $E_{mesons}$ is the sum of the energies of all the mesons.

From the first two panels of this figure, we see that baryons carry almost all of the hadron energy for a vast majority of events when $0 \leq E_{had} \leq 5$ GeV. For larger values of hadron energy, the energy fraction carried by the baryons becomes smaller untill it becomes negligibly small for $E_{had} > 20$ GeV. Therefore, in this section, we study the correlation between the energy carried by the baryons and the hits produced by them in ICAL.
 
\begin{table}[tbp]
\centering
\begin{tabular}{|c|c|c|c|c|c|c|}
\hline
$E_{\rm had}$ (GeV)& hits:0-6 & hits:6-10 & hits:10-20 & hits:20-30 & hits:30-50 & hits:$\geq$ 50 \\
\hline
0-0.5 & 231949 & 8929 & 612 & 41 & 15 & 12 \\ 
\hline
0.5-1 & 79896 & 15699 & 2624 & 33 & 21 & 5 \\
\hline
1-1.5 & 29302 & 8974 & 2869 & 32 & 5 & 6 \\
\hline
1.5-2 & 15570 & 5356 & 2478 & 52 & 3 & 6 \\    
\hline
2-3 & 17551 & 5632 & 3051 & 112 & 10 & 4 \\ 
\hline
3-4 & 9455 & 3031 & 1882 & 83 & 6 & 4 \\
\hline
4-5 & 5756 & 1890 & 1278 & 101 & 10 & 1 \\
\hline
5-7 & 6756 & 2169 & 1569 & 141 & 10 & 0  \\
\hline
7-10 & 4963 & 1582 & 1288 & 195 & 28 & 3 \\
\hline
10-20 & 5529 & 1862 & 1685 & 477 & 156 & 12 \\
\hline
20-30 & 1641 & 544 & 567 &  208 & 153 & 16 \\
\hline
30-50 & 1088 & 369 & 383 & 195 & 165 & 68 \\
\hline
$\geq$50 & 414 & 143 & 167 & 84 & 81 & 67 \\
\hline
\end{tabular}
\caption{Event distribution in "number of baryon hits" for various different values of $E_{had}$.}
\label{table:BaryonhitsEhad}
\end{table}
\subsection{Baryon hits}
In order to do this analysis, we need to know (a) the energy carried by all the baryons ($E_{baryon}$) and (b) the hits produced by these baryons ("baryon\_hits") in each event. It is straight forward to obtain $E_{baryon}$ from NUANCE. To calculate baryon\_hits, we used  the following method: We took all the $\nu_{\mu}$CC events generated by NUANCE and looked at the particle content of all the mesons in these events. We found that essentially all of these mesons are pions and kaons and the heavier mesons form less than $1\%$ of all the mesons. So, we did a Geant4 simulation of these events after turning off the muons, the pions and the kaons. It is expected that the resulting number of hits is essentially due to the baryons produced in these events. The number of events as a function of the number of baryon hits, for various different values of $E_{\rm had}$ are given in table \ref{table:BaryonhitsEhad}.

\begin{figure}[H]
\centering
\includegraphics[width=15cm, height=12cm]{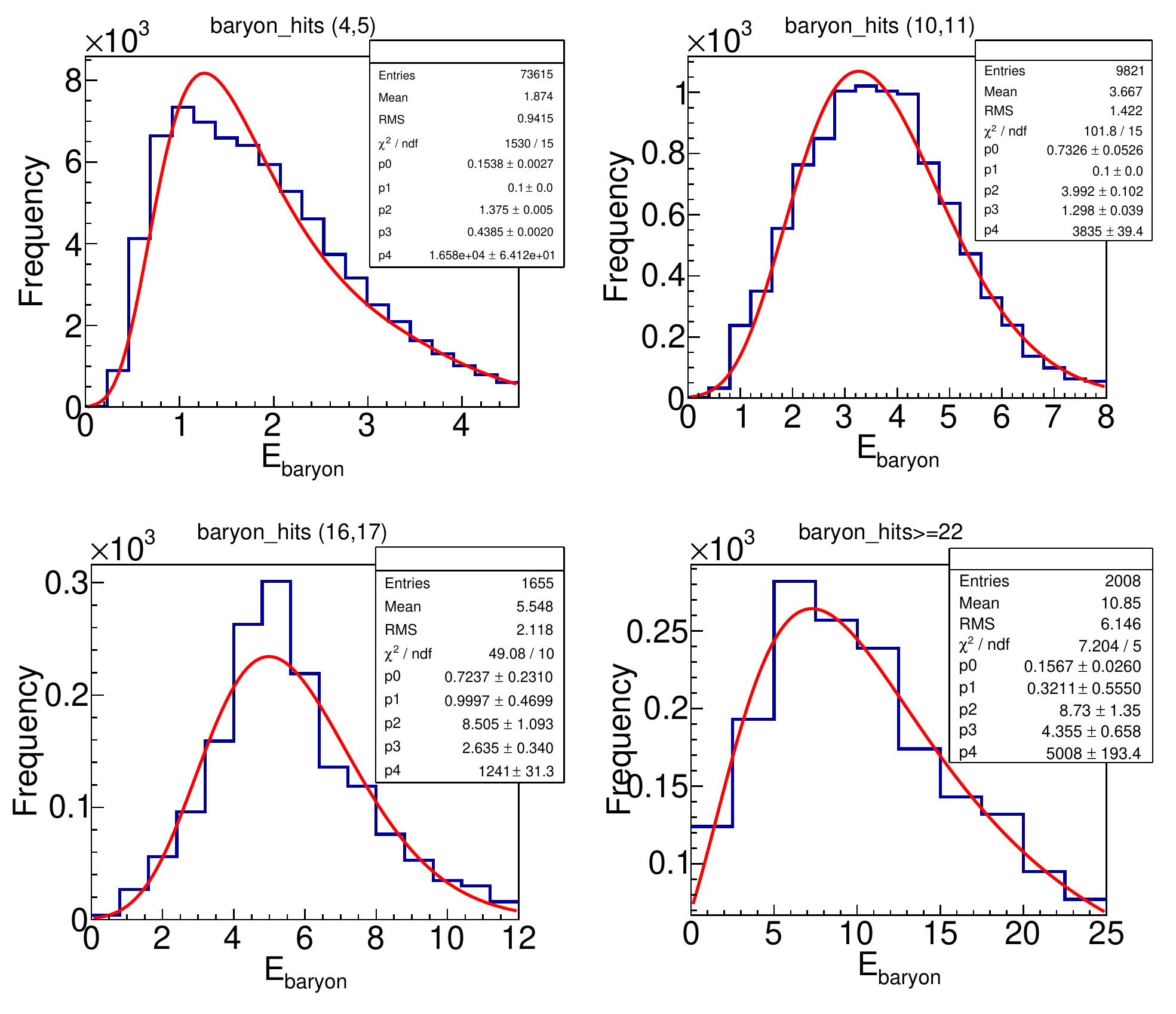}
\caption{$E_{\rm baryon}$ distribution for hit ranges (4,5), (10,11), (16,17) and ($\geq$22)}
\label{fig:Ebary}
\end{figure}

After obtaining $E_{baryon}$ and baryon\_hits for each event, we define a set of ranges of baryon\_hits. For each range we plot the histogram of frequency versus $E_{baryon}$ and fit it with Vavilov distribution \cite{k}. We tested a number different sets of ranges until we found an optimal set of ranges for which the fitted distributions matched the histograms very well. From the  fits done for this optimal set of ranges, we determined the average energy of the baryons $E_{baryon-mean}$ and the associated resolution $\sigma_{Eb}$ for each range. A sample of these fits is shown in figures \ref{fig:Ebary}. The values of $E_{baryon-mean}$ and $\sigma_{Eb}$ for each baryon\_hits range are listed in table \ref{table:baryhitsEbary}. For each distribution, these quantities are defined by 
\begin{eqnarray}
E_{\rm baryon-mean}&=&(\gamma - 1 - lnP_0 - P_1)P_3 + P_2, \nonumber \\
\sigma_{\rm Eb}&=&\sqrt{\dfrac{(2 - P_1)}{2P_0}P^2_3},
\label{vavilovdef}
\end{eqnarray}
where "$\gamma$" is the Euler's constant and "$P_i$" are the parameters of the Vavilov fit.

\begin{table}[tbp]
\centering
\begin{tabular}{|c|c|c|c| }
\hline
S. No. & baryon$\_$hits & $E_{\rm baryon-mean}$ & $\sigma_{\rm Eb}$ \\
\hline
1 & 0-1 & 0.898 & 0.582\\
\hline
2 & 2-3 & 1.340 & 0.845\\
\hline
3 & 4-5 & 1.967 & 1.090\\
\hline
4 & 6-7 & 2.569 & 1.237\\
\hline
5 & 8-9 & 3.179 & 1.369\\     
\hline
6 & 10-11 & 3.717 & 1.477\\
\hline
7 & 12-13 & 4.299 & 1.607\\
\hline
8 & 14-15 &  4.837 & 1.774\\
\hline
9 & 16-17 & 5.609 & 2.190\\
\hline
10 & 18-21 & 6.603 & 2.855\\
\hline
11 & $\geq$ 22 & 13.561 & 10.079\\
\hline
\end{tabular}
\caption{baryon$\_$hits and $E_{baryon}$ table.}
\label{table:baryhitsEbary}
\end{table} 

\begin{figure}[H]
\centering
\includegraphics[width=100mm,scale=0.5]{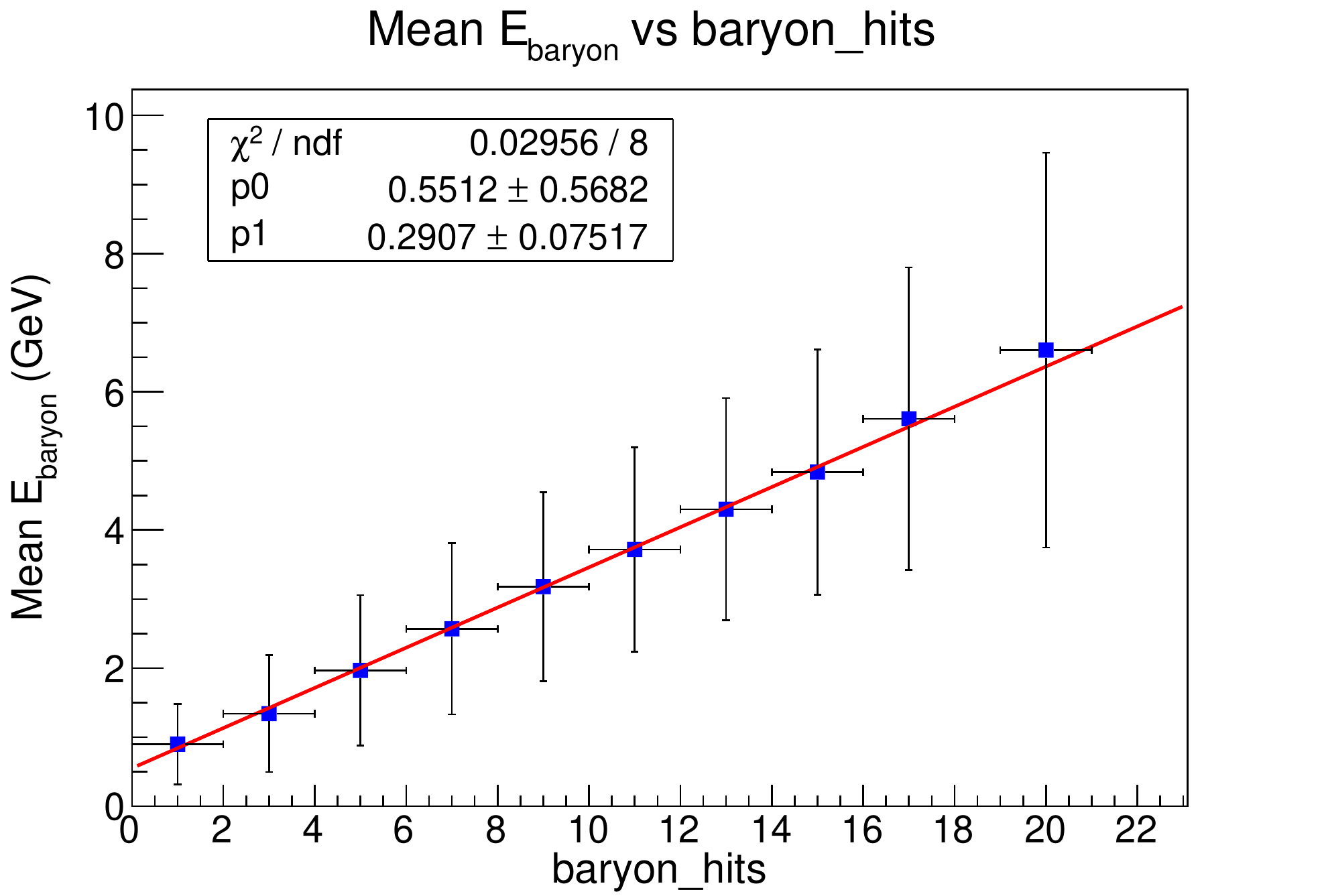}
\caption{$E_{\rm baryon-mean}$ vs baryon \_hits}
\label{fig:baryhitsvsEbary}
\end{figure}

The relation between baryon\_hits and $E_{\rm baryon-mean}$ is shown in figure~ \ref{fig:baryhitsvsEbary}. This relation is well described by the linear fit

\begin{equation}
E_{\rm baryon-mean} =  0.55~(\rm baryon\_hits) + 0.29.
\label{Ebaryon}
\end{equation}

\begin{figure}[H]
\centering
\includegraphics[width=100mm,scale=0.5]{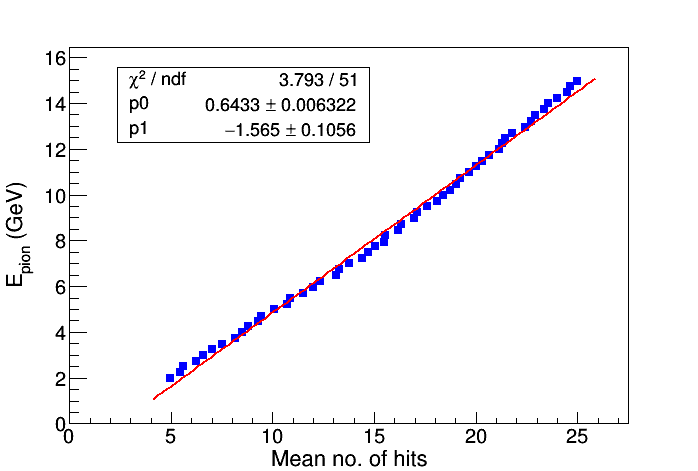}
\caption{$E_{\rm pion}$ vs "Mean no. of hits" for 5.60cm thick iron plates of ICAL}
\label{fig:meanhitsEhad56}
\end{figure}

We compare this relation with the relation obtained from the simulation of isolated pions. This latter simulation was done in reference \cite{g}. The  left panel of figure 2 of this reference gives the plots of "Mean no. of hits" versus $E_{\rm pion}$ for various different thickness values of the iron plates of ICAL. The Geant4 simulation used in our analysis set the iron plate thickness to be $5.6$ cm. We took the data for this thickness from figure 2 of reference \cite{g}
and replotted it in the form $E_{\rm pion}$ versus Mean no. of hits in figure~\ref{fig:meanhitsEhad56}.

We see that this data also shows a linear relationship between these two variables and a linear fit gives the relation 
\begin{equation}
E_{\rm pion}=0.64~(\rm Mean\_no.\_hits) -1.57.
\label{Epion}
\end{equation}
Eventhough there is a linear relation between the no. of hits and the hadron energy for both baryons and pions, as seen from  equations \ref{Ebaryon} and \ref{Epion}, the slope of the pion fit is close to the slope of the baryon fit but the intercepts in the two cases are very different. Therefore, we argue that a proper estimation of hadron energy in atmospheric neutrino events requires doing a full Geant4 simulation of these events and establishing a relation between the number of hits produced by the hadrons and the hadron energy \cite{e}.

\section{Hadron hit analysis}
In this section, we analyse the hits generated by the hadrons produced in the $\nu_\mu$ CC events of atmospheric neutrinos. We establish a correlation between these hadron hits and the energy of the hadrons. We will also calculate the energy resolution for each given hadron energy.  

\subsection{Hadron hit bank}
The Geant4 simulation of atmospheric neutrino event gives the full information on hits due to all the charged particles. This information, in particular, contains the (X,Y,Z) co-ordinates. The full set of this data is called the "hit bank". The ICAL reconstruction code looks to form a track with a subset of the total number of hits. If a track is not constructed, the event is not processed any further. In our analysis, we do not consider these events for which no track is reconstructed. If a track is constructed, it is identified with a muon and its energy and direction are calculated. If more than than one track is constructed, the longest track is identified with a muon and its energy and direction are taken to be the muon kinematic variables. Once the muon track is identified, the hits which make up such a track are removed from the hit bank information. The remaining hits form \textbf{hadron hit bank}. In the analysis below, the events with no hadron hits are discarded. The hadron hit bank is likely to contain ghost hits. The problem of ghost hits is avoided using the procedure described in the introduction. The resultant hits, after implementation of this procedure, are defined to be "hadron\_hits". The difference between the neutrino energy and the muon energy for a given event, obtained from NUANCE, is defined to be "$E_{had}$" of that event.
\begin{figure}[H]
\centering
\includegraphics[width=15cm, height=8cm]{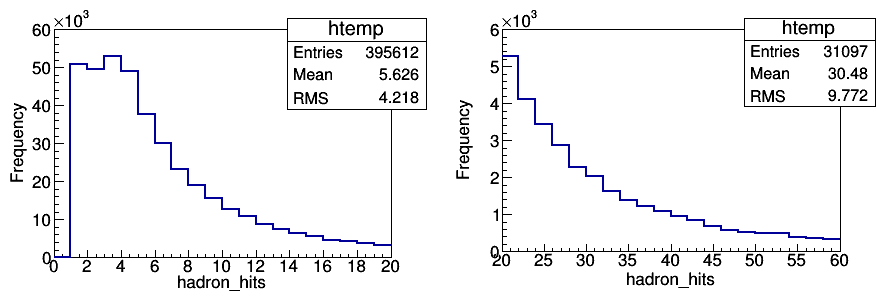}
\caption{ The left (right) panel shows the distribution for the number of hits $<20 (\geq 20)$.}
\label{fig:hithad1}
\end{figure}
The hit distributions of this hit bank are given in figure 
\ref{fig:hithad1}. 
From the hit distribution, we see that most of the events have less than five hits and a vast majority have less than 10 hits. In the following analysis, we will establish a correlation between the number of hits and the hadron energy.
We first divided the event sample into different bins, each with it's own range in the number of hits. For each bin, we plotted the histogram of frequency versus $E_{\rm had}$ and fit it to a Vavilov distrbution. Various different hit ranges were tried out and the procedure was repeated till we obtained an optimal set of hit ranges for which the Vavilov distribution provided a good fit for each of the frequency versus $E_{\rm had}$ histograms. A sample of these histograms, along with the fitted Vavilov distributions, are shown in figures \ref{fig:Ehad}. 
For the bin with (2,3) hits, the Vavilov distribution was not a good fit. Therefore this bin was not used in further analysis. Moreover, such low number of hits may also occur due to just noise. For these reasons, we drop this bin. The hadron hit ranges used and the corresponding Vavilov fit values are shown in table~\ref{table:hitsEhad}. 
\begin{figure}[H]
\centering
\includegraphics[width=15cm, height=22cm]{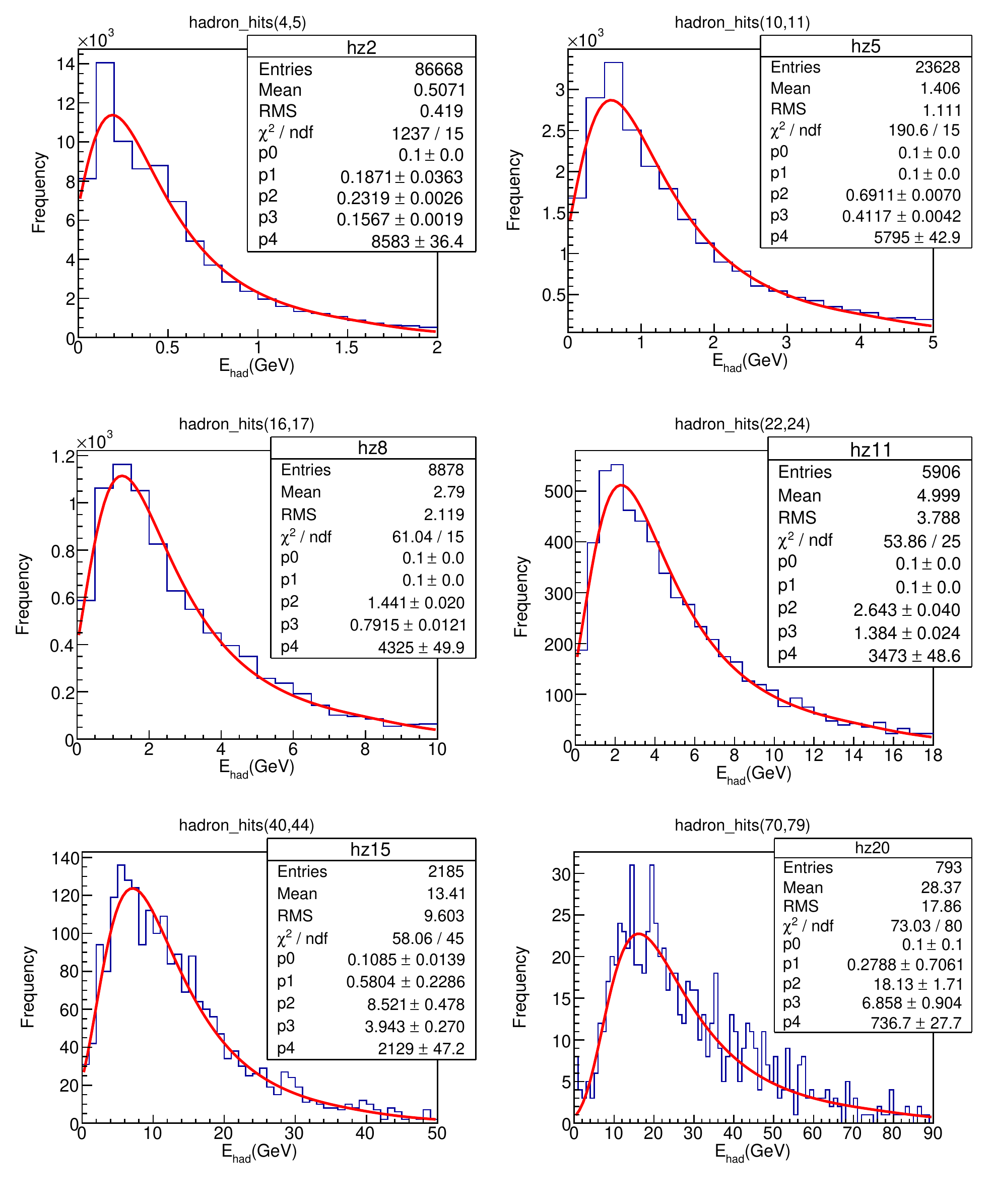}
\caption[$E_{\rm had}$ distribution for hit ranges (4,5), (10,11), (16,17), (22,24),  (40,44) and (70,79).]{$E_{\rm had}$ distribution for hit ranges (4,5), (10,11), (16,17), (22,24), (40,44) and (70,79).}
\label{fig:Ehad}
\end{figure}
\begin{table}[tbp]
\centering
\begin{tabular}{|c|c|c|c|}
\hline
S. No. & hadron$\_$hits & $E_{\rm had-mean}$ & $\sigma_{\rm Eh}$ \\
\hline
1 & 2-3 & 0.356 & 0.356\\ 
\hline
2 & 4-5 & 0.497 & 0.472\\
\hline
3 & 6-7 & 0.939 & 1.111\\
\hline
4 & 8-9 & 1.220 & 1.250\\        
\hline
5 & 10-11 & 1.424 & 1.269\\ 
\hline
6 & 12-13 & 1.892 & 1.687\\
\hline
7 & 14-15 & 2.315 & 1.976\\
\hline
8 & 16-17 & 2.850 & 2.440\\
\hline
9 & 18-19 & 3.639 & 3.094\\
\hline
10 & 20-21 & 4.130 & 3.367\\
\hline
11 & 22-24 & 5.106 & 4.266\\
\hline
12 & 25-29 & 6.572 & 5.411\\
\hline
13 & 30-34 & 8.391 & 6.718\\
\hline
14 & 35-39 & 11.397 & 9.069\\
\hline
15 & 40-44 & 13.320 & 10.083\\
\hline
16 & 45-49 & 15.311 & 10.891\\
\hline
17 & 50-54 & 16.694 & 11.586\\
\hline
18 & 55-59 & 20.203 & 14.196\\
\hline
19 & 60-69 & 24.510 & 16.834\\
\hline
20 & 70-79 & 29.105 & 20.119\\
\hline
21 & 80-99 & 32.444 & 21.135\\ 
\hline
22 & $\geq$ 100 & 44.4066 & 24.679\\
\hline
\end{tabular}
\caption{The ranges of hadron hits and the corresponding fit values of Vavilov distributions.}
\label{table:hitsEhad}
\end{table}
\begin{figure}[H]
\centering
\includegraphics[width=110mm,scale=0.5]{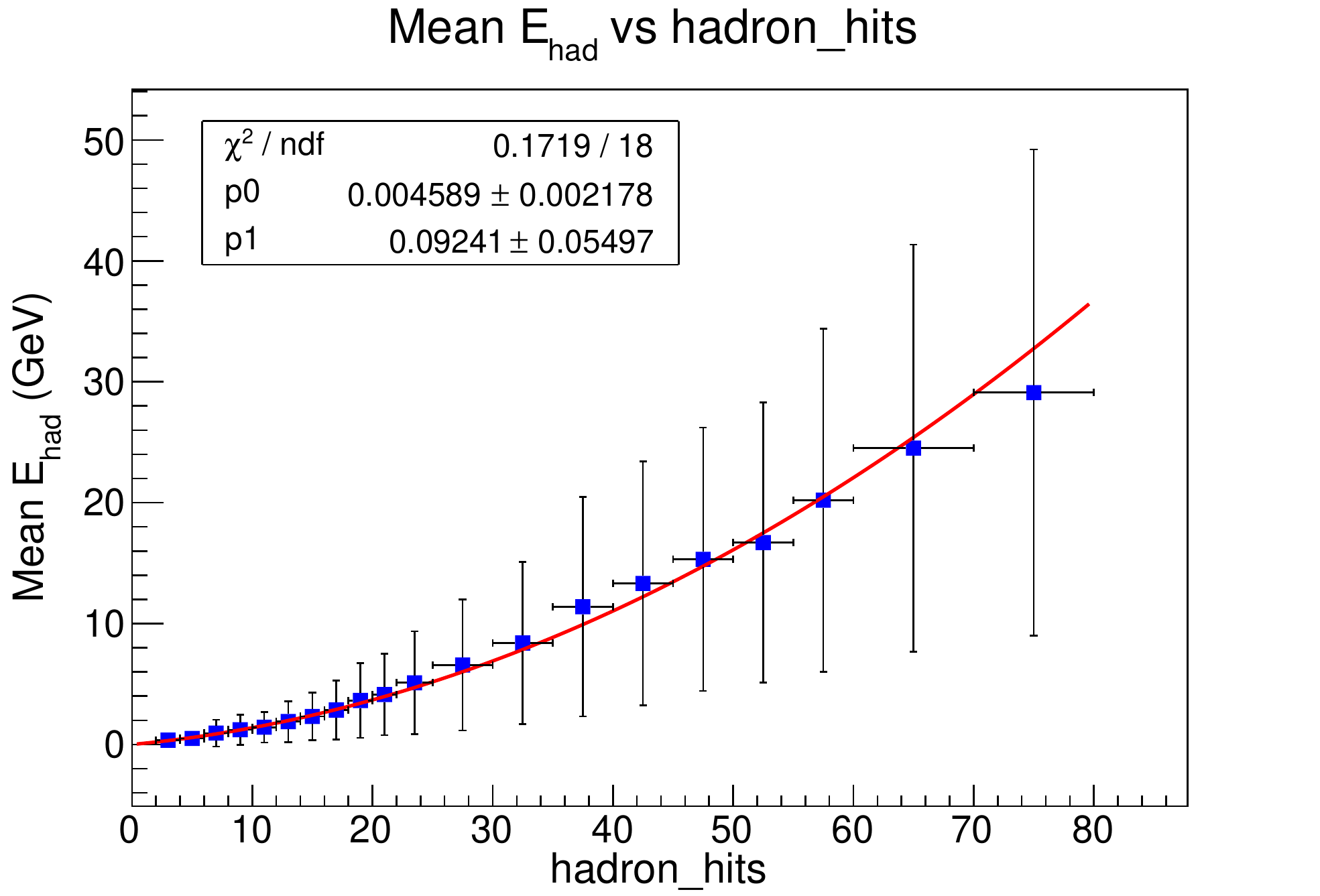}
\caption{$E_{\rm had-mean}$ vs hadron \_hits}
\label{fig:hadronhitsvsEhad}
\end{figure} 
The values of $E_{\rm had-mean}$ and $\sigma_{Eh}$ are calculated for each distribution from the corresponding Vailov fit parameters $P_i$, using equations similar to those in eq.~(\ref{vavilovdef}). 
We plotted hadron hits versus $E_{\rm had-mean}$ in figure \ref{fig:hadronhitsvsEhad}. When this plot was fitted with a linear function, the estimate of hadron energy was too low for hadron hits $>40$. Therefore we did a fit with a quadratic function and obtained 
\begin{equation}
E_{\rm had-mean} \simeq 0.09x + 0.005x^{2}, 
\end{equation}
where $x$ represents the number of hadron hits.
We also plotted $\sigma_{Eh}^2$ versus $E_{\rm had-mean}$ in figure \ref{fig:sigmaEhvsEhad}. This energy resolution is parametrized as $\sigma(E)/E = \sqrt{a^2/E + b^2}$ \cite{g}. A fit to the plot gives the values 
\begin{equation}
\left(\frac{\sigma_{\rm Eh}}{E_{\rm had-mean}}\right) = \sqrt{\frac{2.1 \pm 0.4}{E_{\rm had-mean}} + 0.38 \pm 0.3},
\end{equation} 
leading to $a=1.45 \pm 0.14$ and $b=0.62 \pm 0.02$.
\begin{figure}[H]
\centering
\includegraphics[width=12cm, height=9.1cm]{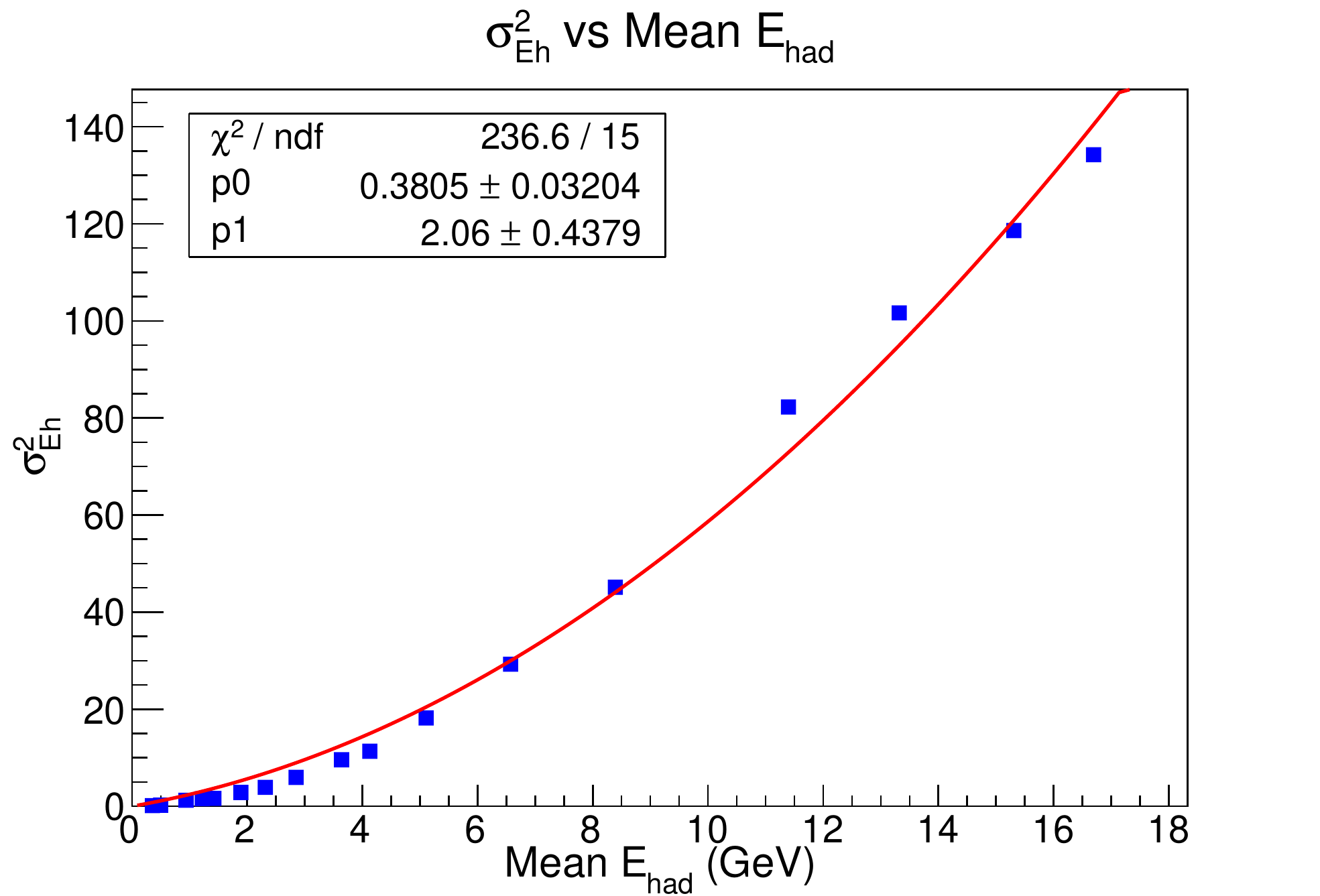}
\caption{$\sigma_{\rm Eh}^2$ vs $E_{\rm had-mean}$}
\label{fig:sigmaEhvsEhad}
\end{figure}
We believe that $E_{\rm had-mean}$ and $\sigma_{\rm Eh}$ obtained in this section form the correct representation of hadron energy and its resolution in atmospheric neutrino events.

In a previous work, the authors of ref.\cite{e} also have used the hadron hit information from the Geant4 simulation of NUANCE generated atmospheric neutrino events. There are a number of differences in the procedure they used and in the procedure used in this work. 

\begin{itemize}
\item Their data set consists of 1000 years of atmospheric neutrino events, whereas our  set consists of 100 years of data.  
\item They obtained hadron hit bank information by doing the Geant4 simulation of an event with the muon turned off at the input level. In our case, we did the full Geant4 simulation of all the charged particles in the event and subtracted the hits which went into the track reconstrcution. This is the procedure which will be utilized in the case of actual data.
\item The avalanche produced in an RPC by one charged particle can, quite often, produce hits in two adjacent strips. Thus, the number of hadron hits in  an RPC is likely to be larger than the number of charged particles passing through it. This feature is built into Geant4 through the option \textit{multiplicity}. The authors of ref.\cite{e} kept this option \textit{off} and hence obtained a smaller number of hits for a given hadron energy. In our case, we kept the multiplicity option \textit{on} and obtained about 30 to 40\% larger numbner of hits for the same hadron energy. This is a more realistic simulation of the detector.
\end{itemize}
Because of points 2 and 3, the procedure we used is modelled more closely to what happens in the detector.
\section{Conclusion}
In this work, we attempted to obtain an estimate of the energy of hadrons produced in a charged current interaction of an atmospheric muon neutrino/anti-neutrino in ICAL at INO. This was done by doing a full Geant4 simulation of atmospheric neutrino events generated by NUANCE. We have used the un-oscillated data simulated for a period of 100 years. We have established the following results:
\begin{itemize}
\item For $E_{had}$ < 5 GeV, almost all of the hadron energy is carried by the baryons. 
\item The relation between the number of hits and the energy of hadrons is very different for the two cases when the hadrons are mesons and when the hadrons are baryons.
\item When the events are classified into bins with different number of hadron hits, the resulting spectra are reasonably well described by Vavilov distributions. 
\item There is a good correlation between the number of hits and the mean value of $E_{Ehad}$ of the Vavilov distributions. 
\item The width ($\sigma$) of the Vavilov distributions is related to the mean energy ($E_{\rm had-mean}=E$) through the expected relation $\sigma (E)/E=\sqrt{(a^2/E +b^2)}  $.
\end{itemize}
It will be interesting to redo the analysis of the physics sensitivities of ICAL at INO using the hadron energy estimates and resolution functions presented here.
\acknowledgments

We would like to thank Dr. Ali Ajmi who provided valuable help in the initial stages of this work. We also thank Prof. Gobinda Majumder, Dr. Susnata Seth, Prof. Prafulla Behera and Dr. Jafar Sadiq for discussions at different times. Thanks are also due to Prof. Raj Gandhi and Prof. Amol Dighe who gave valuable feedback at various intermediate stages.





\end{document}